\documentclass[preprint,aps,showpacs,prb]{revtex4}

\usepackage{graphicx}
\usepackage{dcolumn}
\usepackage{bm}
\usepackage{epsfig}
\graphicspath{{./}{data_paper/}}
\DeclareGraphicsExtensions{.ps,.ps.gz}
\begin{document}

\title{\bf Critical analysis of a variational method used to describe the 
molecular electron transport}

\author{Ioan B\^aldea}
\email{ioan.baldea@pci.uni-heidelberg.de}
\altaffiliation[Also at ]{National Institute for Lasers, Plasma, 
and Radiation Physics, ISS, POB MG-23, RO 077125 Bucharest, Romania.}
\author{Horst K\"oppel} 
\affiliation{Theoretische Chemie,
Physikalisch-Chemisches Institut, Universit\"{a}t Heidelberg, Im
Neuenheimer Feld 229, D-69120 Heidelberg, Germany}

\date{\today}

\begin{abstract}
In a recent paper [I.\ B\^aldea and H.\ K\"oppel, \prb {\bf 78}, 115315 (2008)], we showed that 
a variational approach [P.\ Delaney and J.\ C.\ Greer, \prl {\bf 93}, 036805 (2004)] 
proposed to compute the electron transport through molecules, which is based on boundary constraints of 
the Wigner function, is unable to correctly describe the zero-bias conductance of the simplest 
uncorrelated and correlated systems. In the present paper, we extend our previous analysis 
of the linear response limit of that approach, by considering, instead of the Wigner function, 
general constraints. We demonstrate that, if, as usual in transport theories, 
the quasi-particle distributions in electrodes are constrained, this method yields the 
completely unphysical result that the zero-bias conductance vanishes.
Therefore, we conclude that the variational approach itself is defective. 
\end{abstract}

\maketitle

\renewcommand{\topfraction}{1}
\renewcommand{\bottomfraction}{1}
\renewcommand{\textfraction}{0}

\section{Introduction}
\label{sec:introduction}
In spite of significant advances, the field of molecular electronics 
continues to remain confronted with problems unsatisfactorily clarified
both from the experimental and the theoretical sides.
One major issue is the large discrepancy, 
often of several orders of magnitude,\cite{Wenzel:02,Nitzan:03,Stokbro:03} 
between the measured currents through molecules and values calculated theoretically.
Open questions in electron transport through nanostructures, which remain unresolved 
after many years of research, appear also in other related areas, e.~g., transport 
through semiconducing quantum dots \cite{baksmaty:136803,NPJ-9-2007}
\par
The simplest theoretical calculations of the transport in nanoscopic/molecular 
systems rely upon the Landauer formalism,\cite{Landauer:70,Buttiker:85} 
which ignores electron correlations. The latter are accounted for in more elaborated 
approaches, like those based on nonequilibrium 
Green functions (NEGFs),\cite{HaugJauho,Datta:05} time-dependent density matrix 
renormalization group (DMRG),\cite{White:04,Daley:04,White:05} and  
numerical renormalization group (NRG).\cite{Bulla:08}. In the most popular 
\emph{ab initio} calculations in molecular electronics, the NEGF method is combined 
with calculations based on the density functional theory (DFT).
\cite{Taylor:01,Brandbyge:02,Rocha:06} 
Other approaches are based on various many-body schemes.
\cite{Wenzel:03,Muralidharan:06,Thygesen:07}
Although the role of uncontrollable experimental factors should not be underestimated 
(compare, for example, the contradictory results for one and the same system 
\cite{Reed:97,Xiao:04,Dadosh:05,tsutsui:06}), the ability to predict currents  
comparable with measured values continues to represent an important challenge 
for theory.
\par
Out of the existing theoretical methods, 
that proposed and used in Refs.\ 
\onlinecite{DelaneyGreer:04a,DelaneyGreer:04b,DelaneyGreer:06,Fagas:06,Fagas:07} 
seemed to be very appealing, 
because it claimed to correctly reproduce currents experimentally measured 
through correlated molecules and yielded a few other plausible results. 
It represents a variational approach 
with constraints imposed on the Wigner function at the device-electrode interfaces. 
The latter aspect is original: in usual transport theories, 
the electron distributions are constrained to be the Fermi functions.
In a recent paper,\cite{Baldea:2008b} we inquired the validity of this method 
within the realm of theory and demonstrated that, for the simplest uncorrelated 
and correlated systems, its results are completely unphysical. However, because our 
critique primarily envisaged the imposed boundary conditions, one might 
still hope that this approach can be remedied by imposing other constraints.
To investigate whether this is the case or not is the main purpose of our present work.
\par
The remaining part of the paper is organized in the following manner.
The variational approach considered here, which generalizes the approach 
proposed in Ref.\ \onlinecite{DelaneyGreer:04a}, will be described in Sect.\ \ref{sec-variational}. 
Its linear response limit will be considered in the next Sect.\ \ref{sec-lra}. 
As an important point of the present work, which is emphasized 
in Sect.\ \ref{sec-Fermi-BC}, we argue that, contrary to what was claimed in 
Ref.\ \onlinecite{DelaneyGreer:04a}, 
strong electron correlations in the device 
do not preclude to impose usual boundary conditions, namely, to constrain 
the single-electron distributions 
in electrodes to be the Fermi distributions. 
The main result deduced in Sect.\ \ref{sec-Fermi-BC} is that, even  
with these usual boundary conditions, 
the variational approach is inappropriate: it unphysically 
predicts a zero-bias conductance that vanishes in situations 
where this is definitely not the case (e.~g., uncorrelated systems, 
Coulomb blockade peaks, Kondo plateau). 
Conclusions are presented in Sect.\ \ref{sec-conclusion}. 
\section{Description of the variational approach}
\label{sec-variational}
Some years ago, Delaney and Greer (DG) proposed a method to calculate the 
steady-state electric transport through correlated molecules 
connected to two electrodes (two-terminal transport set-up) within a 
many-body approach.\cite{DelaneyGreer:04a} It relies upon many-body calculations 
for a finite cluster consisting of the (nano)device (the molecule of interest) 
and (small) parts of electrodes with and without voltage bias. 
Their approach can be summarized as follows.
A constant voltage bias, which is described by a term $W$ in the Hamiltonian, 
drives the cluster at zero temperature ($T=0$) from its ground state $\Psi_{0}$ 
($H\vert\Psi_{0}\rangle = \mathcal{E}_{0}\vert\Psi_{0}\rangle$) 
to a steady state $\Psi$ characterized by a constant current flow.
\par 
(i) The steady state wave function $\Psi$ is determined by 
minimizing the total energy $\mathcal{E}$ in the presence of the applied voltage
\begin{equation}  
\label{eq-total-energy}  
\mathcal{E} = 
\left \langle \Psi \right \vert H \left \vert \Psi \right \rangle
+ \left \langle \Psi \right \vert W \left \vert \Psi \right \rangle 
\end{equation}  
subject to the following constraints:
\par
(ii) for the electrons flowing from electrodes into the device, 
the Wigner function at the two device-electrode contacts determined from $\Psi$ 
in the presence of the voltage bias be equal to that obtained from the 
many-body ground state $\Psi_{0}$ of the cluster in the absence of the bias;
\par
(iii) if not automatically satisfied,\cite{DelaneyGreer:04a,DelaneyGreer:04b} 
the steady-state electric current $J$ should be constrained to be position 
($q$) independent
\begin{equation}
\label{eq-j-average}    
\left\langle \Psi \right \vert j_{q}\left \vert \Psi \right \rangle = J ,
\end{equation}  
as required by the equation of continuity 
($\partial \rho/\partial t + \mathbf{\nabla} \raisebox{0.7ex}{.}  \mathbf{j} = 0 $);
\par
(iv) the steady-state wave function is normalized
\begin{equation}  
\label{eq-norm-Psi}
\left\langle \Psi \right. \left \vert \Psi \right \rangle = 1.
\end{equation} 
\par
In our recent work (Ref.\ \onlinecite{Baldea:2008b}), 
we showed that this approach is incorrect, 
as it fails to reproduce well-established results both for 
uncorrelated and correlated systems. 
The failure for the uncorrelated case is important for two reasons.
First, in this case very large systems can be treated by this method without 
any other approximations than the method itself. 
This demonstrates that 
the completely unphysical results it generates are due to the 
method \emph{itself} and cannot be assigned to the fact that, realistically, 
only very small parts 
of electrodes can be accounted for within accurate many-body calculations.
Second, it challenges the need for and/or the usefulness of the Wigner function.
Imposing boundary conditions by means of the 
Wigner function is only \emph{one} ingredient of this method, and one may 
ask whether employing other boundary conditions 
could mender this variational approach.
\par
In the present paper, we shall consider such alternative choices.
For the moment, we do not specify the properties to be constrained. 
Rather, instead of the above condition (ii), we shall consider general constraints of the form
\begin{equation}  
\label{eq-Q}  
\left \langle \Psi \right \vert Q_{\kappa} \left \vert \Psi \right \rangle 
= 
\left \langle \Psi_{0} \right \vert Q_{\kappa} \left \vert \Psi_0 \right \rangle 
\end{equation}  
for a set of Hermitian operators $\left\{Q_{\kappa}\right\}$.\cite{hermitian}
Condition (ii) represents a particular case of Eq.\ (\ref{eq-Q}), where 
$Q_{\kappa}$ is chosen to be the Fano operator \cite{Baldea:2008b}
\begin{equation}
F(x, p) = \sum_{l, \sigma} a^{\dagger}_{x-l,\sigma} a_{x+l,\sigma} 
e^{-2 i p l/\hbar} ,
\label{eq-fano}  
\end{equation}  
with $a_{x,\sigma}$ and $a_{x,\sigma}^{\dagger}$ denoting the annihilation and creation 
operators for an electron of spin $\sigma$ located at site $x$.
\par
For the sake of simplicity, 
we shall consider one-dimensional discrete systems, with 
left ($L$) and right ($R$) electrodes containing electrons described within 
the tight-binding nearest-neighbor approximation ($e=L,R$). 
Although more general cases can also be considered, we shall assume 
a total Hamiltonian of the form
\begin{eqnarray}  
H & = & H_{L} + H_{R} + H_{D} + H_{D,e}\nonumber \\
H_{L} & = & \mu_{L} \sum_{l\leq q_{L}, \sigma}  a_{l,\sigma}^{\dagger} a_{l,\sigma}^{}
-  \sum_{l \leq q_{L},\sigma} t_{l,\sigma}\left(
  a_{l,\sigma}^{\dagger} a_{l-1,\sigma}^{} 
+ a_{l-1,\sigma}^{\dagger} a_{l,\sigma}^{}
\right) , \nonumber \\ 
H_{R} & = &
 \mu_{R} \sum_{r \geq q_{R},\sigma}  a_{r,\sigma}^{\dagger} a_{r,\sigma}^{}
        - \sum_{r \geq q_{R},\sigma} t_{r,\sigma} \left( 
  a_{r,\sigma}^{\dagger} a_{r+1,\sigma}^{} 
+ a_{r+1,\sigma}^{\dagger} a_{r,\sigma}^{}\right) , \label{eq-hamiltonian}  \\
H_{D,e} & = &
- \sum_{\sigma} t_{q_{L},\sigma}\left(a_{q_{L},\sigma}^{\dagger} a_{q_{L}+1,\sigma}^{} + a_{q_{L}+1,\sigma}^{\dagger} a_{q_{L},\sigma}^{}\right)
- \sum_{\sigma} t_{q_{R},\sigma}\left(a_{q_{R}-1,\sigma}^{\dagger} a_{q_{R},\sigma}^{} + a_{q_{R},\sigma}^{\dagger} a_{q_{R}-1,\sigma}^{}\right)
\nonumber \\
H_{D} & = &
\sum_{\sigma} \sum_{q=q_{L}+1}^{q_{R}-1}\varepsilon_{q} a_{q, \sigma}^{\dagger} a_{q, \sigma}^{}
- \sum_{\sigma} \sum_{q=q_{L}+1}^{q_{R}-2} t_{q,\sigma} \left( a_{q, \sigma}^{\dagger} a_{q+1, \sigma}^{}
+ a_{q+1, \sigma}^{\dagger} a_{q, \sigma}^{} \right)
+  H_{int}\left( n_{q_{L}+1,\sigma}, \ldots ,n_{q_{R}-1,\sigma}\right) . \nonumber
\end{eqnarray}  
Above, $a_{l,\sigma}, a_{r,\sigma}$, and $a_{q,\sigma}$ 
($a_{l,\sigma}^\dagger, a_{r,\sigma}^\dagger$, and $a_{q,\sigma}^\dagger$) 
represent annihilation (creation) operators of electrons with spin $\sigma$ at sites 
$l$, $r$, and $q$ belonging to the left (L) and right (R) electrodes, 
and to the device (D), respectively. $\mu_{L,R}$ denote electrode chemical potentials, 
$t$'s are hopping integrals, and $\varepsilon$'s are on-site energies in the device.
The above model incorporates 
a rather general electron-electron interaction $H_{int}$ in the device. We need not 
to specify an explicit form of $H_{int}$, we solely assume that it can be expressed in terms of the 
electron numbers $n_{q,\sigma} \equiv a_{q,\sigma}^{\dagger} a_{q,\sigma} $.
For concreteness, one can consider a particular explicit form $H_{int} = \sum_{\sigma_1,\sigma_2}
\sum_{q_L\leq q_1,q_2 \leq q_R}
U_{q_1,\sigma_1,q_2,\sigma_2} n_{q_1,\sigma_1} n_{q_2,\sigma_2}$. 
Important particular cases thereof are multi-site nanodevices  
described by extended Hubbard models
($U_{q_1,\sigma_1,q_2,\sigma_2} = U\delta_{q_1,q_2}\delta_{\sigma_1,-\sigma_2}$, 
$U_{q_1,\sigma_1,q_2,\sigma_2} = V\delta_{q_1,q_2\pm 1}$), or devices consisting of a single site ($q_D$),
like point contacts or single-electron transistors (Anderson impurity model), which amounts
to choose $U_{q_D,\sigma_1,q_D,\sigma_2} = U \delta_{\sigma_1,-\sigma_2}$, where $U=0$ or $U\neq 0$, respectively. 
One should emphasize that the model specified above includes both uncorrelated 
($U_{q_1,\sigma_1,q_2,\sigma_2}\equiv 0$) and strongly correlated cases. 
Phenomena like the Coulomb blockade or 
the Kondo effect described by the Anderson impurity model are typical examples 
of strong electron correlations, where the single-particle picture completely breaks down. 
\par
Although not explicitly needed for the subsequent considerations, we give 
for illustration the form of the term $W$ pertaining to the applied bias $V$
\begin{equation}
\label{eq-W}  
W =   \frac{V}{2} \sum_{l\leq q_{L}, \sigma}  a_{l,\sigma}^{\dagger} a_{l,\sigma}^{}
+ \sum_{\sigma} \sum_{q=q_{L}+1}^{q_{R}-1}V_{q} a_{q, \sigma}^{\dagger} a_{q, \sigma}^{}
- \frac{V}{2} \sum_{r \geq q_{R},\sigma}  a_{r,\sigma}^{\dagger} a_{r,\sigma}^{}  ,
\end{equation}
where the concrete way how the potential $V_q$ drops across the device 
($ - V/2 \leq V_{q} \leq V/2$) is not needed.
\par
For model (\ref{eq-hamiltonian}), the electric current operator  $j_{q}$ 
at site $q$ ($e=\hbar = 1$) has the expression \cite{Caroli:71}
\begin{equation}
\label{eq-j}  
j_{q} = i \sum_{\sigma} t_{q,\sigma} \left(
a^{\dagger}_{q,\sigma} a_{q+1,\sigma} - a^{\dagger}_{q+1,\sigma} a_{q,\sigma} 
\right) .
\end{equation}
If the total cluster possesses $N$ sites, there are $N-1$ values 
$\langle \Psi \vert j_{q} \vert \Psi \rangle$. Therefore, the equation of continuity, 
Eq.\ (\ref{eq-j-average}), yields $N-2$ constraints of the form
\begin{equation}
\label{eq-j-constraints}  
\left\langle \Psi \right \vert j_{q} - j_{q_0}\left \vert \Psi \right \rangle = 0,
\end{equation}
where $q_0$ stands for a fixed arbitrary site.
For the moment, we need not to specify the $Q_{\kappa}$'s in Eqs.\ (\ref{eq-Q}). 
\par
To simplify the analysis, we shall assume that $\Psi_{0}$ is nondegenerate.
The wave function $\Psi$ will be expanded in terms of the complete set 
of orthonormalized eigenstates of $H$ ($H\vert\Psi_{n}\rangle = E_{n}\vert\Psi_{n}\rangle$). 
\begin{equation}  
\left\vert \Psi \right \rangle 
=  A_{0} \left \vert \Psi_{0} \right \rangle + 
\sum_{n\neq 0} A_{n} \left \vert \Psi_{n} \right \rangle .
\label{eq-Psi}  
\end{equation}  
To simplify the discussion, we shall suppose that the model parameters entering 
Eq.\ (\ref{eq-hamiltonian}) are real. This enables us to choose the eigenstates $\Psi_{n}$ as  
real. However, the expansion coefficients 
$A_{n}$ can still be complex, and thus 
the general boundary conditions (\ref{eq-Q}) and the equation of continuity (\ref{eq-j-average}) 
can be satisfied. We shall suppose that $A_{0}$ is real, which 
amounts to choose the phase factor appropriately.
\par
Starting with a system in the ground state $\Psi_{0}$, we shall look for the solution
$\Psi$, which minimizes the quantity
\begin{equation}
\label{eq-min}  
\left \langle \Psi \right \vert H \left \vert \Psi \right \rangle
+ \left \langle \Psi \right \vert W \left \vert \Psi \right \rangle 
- \omega \langle \Psi \vert \Psi \rangle
- \sum_{\kappa} \lambda_{\kappa} 
\left(
\left \langle \Psi \right \vert Q_{\kappa} \left \vert \Psi \right \rangle 
-
\left \langle \Psi_{0} \right \vert Q_{\kappa} \left \vert \Psi_0 \right \rangle 
\right)
+ i \sum_{q \neq q_0} \chi_{q}
\left\langle \Psi \right \vert j_{q} - j_{q_0}\left \vert \Psi \right \rangle
\end{equation}
with the supplementary constraints (\ref{eq-norm-Psi}), (\ref{eq-Q}), and 
(\ref{eq-j-constraints}). This yields a set of equations, 
wherefrom the optimum values of 
the coefficients $A_{0}$ and $A_{n}$, and the (real) Lagrange multipliers 
$\omega$, $\chi_{q}$, and $\lambda_{\kappa}$ can be determined.
\section{Linear response approximation}
\label{sec-lra}
Similar to Ref.\ \onlinecite{Baldea:2008b}, 
we shall work out the linear response approximation
of the method described above, which should enable us to compute 
the zero-bias conductivity.
Therefore, we shall only consider changes to the relevant 
quantities of the order $\mathcal{O}(W)$.
\par
While in general this minimization represents a 
difficult nonlinear problem, it considerably simplifies in the linear response limit,
applicable for a small applied bias.
In this limit, the minimization amounts to solve a linear system of equations, 
which possesses a unique solution $A_{0} = 1 + \mathcal{O}(W^2)$, 
$A_{n} = \mathcal{O}(W)$ for $n\neq 0$, $\lambda_{\kappa} = \mathcal{O}(W)$,
$\omega = E_{0} + \mathcal{O}(W)$, 
and $\chi_{q} = \mathcal{O}(W)$. 
\par
The quantities entering the minimization problem within the linear response approximations are
\begin{eqnarray}
\label{eq-Q-n}  
& & 
\left\langle \Psi_{n} \right \vert Q_{\kappa} \left \vert \Psi_{0} \right \rangle 
\equiv Q_{\kappa, n} \equiv \mathcal{X}_{\kappa,n} + i \mathcal{Y}_{\kappa,n} , \\
\label{eq-W-n}  
& & \left\langle \Psi_n \left \vert W \right \vert \Psi_0 \right \rangle  \equiv \mathcal{W}_{n} , \\
\label{eq-J-n}  
& & 
\left\langle \Psi_n \right \vert j_{q} - j_{q_{0}}\left \vert \Psi_0 \right \rangle \equiv  i \mathcal{J}_{q,n} .
\end{eqnarray}  
\par
Above and throughout, the calligraphic symbols denote real matrix elements.
The minimization yields the following results in the linear response 
approximation. The expansion coefficients are expressed by 
\begin{equation}  
A_{n} = \frac{1}{E_{n} - E_{0}}
\left[
-\mathcal{W}_{n} + 
\sum_{\kappa} \lambda_{\kappa} Q_{q,n}
+i \sum_{q \neq q_{0}} \chi_{q} \mathcal{J}_{q,n}
\right] .
\label{eq-A}  
\end{equation}  
The normalization condition (\ref{eq-norm-Psi}) leads to $\omega = E_{0}$,
and the constraints (\ref{eq-j-constraints}) and (\ref{eq-Q}) yield  
\begin{eqnarray}
\label{eq-Q-constraint}  
& \sum_{n\neq 0} & \left( A_{n}^{\ast} Q_{\kappa,n} + A_{n} Q_{\kappa,n}^{\ast} \right) = 0,\\
\label{eq-J-constraint}  
& \sum_{n\neq 0} & \left( A_{n} - A_{n}^{\ast} \right) \mathcal{J}_{q,n} = 0 .
\end{eqnarray}  
At this point, it is useful to separate the real and imaginary parts of Eqs.\ 
(\ref{eq-A}), (\ref{eq-Q-constraint}), and (\ref{eq-J-constraint}). 
This immediately leads to
\begin{eqnarray}
\displaystyle
\label{eq-lambda}  
& & \sum_{\kappa^\prime} \lambda_{\kappa^{\prime}} 
\sum_{n\neq 0} 
\frac{\mathcal{X}_{\kappa^{\prime},n} \mathcal{X}_{\kappa,n} + \mathcal{Y}_{\kappa^{\prime},n} \mathcal{Y}_{\kappa,n} }{E_{n} - E_{0}}
+ \sum_{q} \chi_{q} \sum_{n\neq 0} 
\frac{\mathcal{Y}_{\kappa,n}  \mathcal{I}_{q, n} }{E_{n} - E_{0}}
=
\sum_{n\neq 0} \frac{\mathcal{W}_{n} }{E_{n} - E_{0}} \mathcal{J}_{q,n} , \\
\label{eq-chi}  
& & 
\sum_{q^\prime} \chi_{q^\prime} \sum_{n\neq 0} \frac{\mathcal{J}_{q^\prime, n} \mathcal{J}_{q, n} }{E_{n} - E_{0}}
+ \sum_{\kappa^\prime} \lambda_{\kappa^{\prime}} 
\sum_{n\neq 0} 
\frac{\mathcal{Y}_{\kappa^{\prime},n} \mathcal{I}_{q^\prime, n} }{E_{n} - E_{0}}
= 0 .
\end{eqnarray}  
Let us assume that 
there are $N_L$ and $N_R$ constraints of the form (\ref{eq-Q}) imposed for the left and 
right electrodes, respectively. 
Eqs.\ (\ref{eq-lambda}) and (\ref{eq-chi}) represent a linear set of 
$N_L + N_R + N - 2$ equations.
Except for accidental cases where the determinants vanish,\cite{nonvanishing-determinants}
these equations determine the $N_L + N_R$ values $\lambda_{\kappa}$ 
and the $N-2$ values $\chi_{q}$ ($q \neq q_{0}$) of the Lagrange multipliers uniquely. 
Once they are known, the expansion coefficients $A_{n}$ can be obtained from 
Eq.\ (\ref{eq-A}), which, in turn, allow to determine the $q$-independent 
steady state current $J = J_{q}$ as
\begin{equation}
\label{eq-J}  
J = -2 i \sum_{\sigma, n\neq 0} 
\left(
\sum_{\kappa} \lambda_{\kappa} \mathcal{Y}_{\kappa, n} + 
\sum_{q \neq q_{0}} \chi_{q} \mathcal{J}_{q^\prime, n}
\right)
\left \langle \Psi_{n} \left \vert j_{q,\sigma} \right \vert \Psi_{0}\right \rangle .
\end{equation}
Because the angular parenthesis in the r.h.s.\ of the above equation is purely imaginary 
[see Eq.\ (\ref{eq-j})], the quantity $J$ is real.
Notice that $\mathcal{W}_{u}$ enters linearly Eqs.\ 
(\ref{eq-lambda}) and (\ref{eq-chi}), and therefore the current computed 
from Eq.\ (\ref{eq-J}) is proportional to the applied bias $V$.
This means that the above minimization procedure 
yields the solution corresponding to the linear response limit.
\par
We end the part devoted to general considerations by addressing a technical 
issue. From the perspective of 
Ref.\ \onlinecite{DelaneyGreer:04a}, the variational approach discussed 
here would be ultimately intended to be used in conjunction with ab initio 
quantum chemical calculations for a real system,
which comprises the device and parts of electrodes.
It would be desirable that the latter are sufficiently large, such that 
the bulk electrode properties are approached. In practice, the size ($N$) 
of the system that can be investigated is inherently finite, and therefore an $N$-dependence of 
the results is unavoidable [cf.\ Eq.~(\ref{eq-j-constraints})].
The best one can hope in a realistic ab initio calculation is to be able to 
increase $N$ until results converge, in a way similar to the much less demanding 
time-dependent DMRG calculations.\cite{White:04,Daley:04,White:05}
In fact, increasing the size is 
so prohibitive that the most ambitious ab initio calculations 
can at most include a few electrode layers in the cluster used for transport calculations, and the 
saturation with increasing $N$ cannot be systematically checked.\cite{grid}
Similarly, the expansion (\ref{eq-Psi}) 
cannot exhaust the multi-electronic Hilbert space of a real system, which 
is infinitely dimensional. 
What one has to do there is to increase the number $\mathcal{N}$ 
of multi-electronic wave functions $\{\Psi_n\}_{n=1}^{\mathcal{N}}$ until reaching convergence.
Of course, the latter limitation only applies to a real system.  
For the discrete cases described by Eq.\ (\ref{eq-hamiltonian}), $\mathcal{N}$ 
rapidly grows with $N$ and can become very large,\cite{finite-N} 
but it remains finite for any finite cluster.
\section{Constraints imposed to the single-electron distributions}
\label{sec-Fermi-BC}
So far, we did not specify the boundary conditions (\ref{eq-Q}).
In Ref.\ \onlinecite{DelaneyGreer:04a}, resorting to the Wigner function 
(in a theoretical approach of the transport in correlated systems) 
was motivated by the fact that 
working with a correlated many-body wave function $\Psi$ 
precludes a description in terms of wave functions and energies 
$\epsilon_i$ for independent electrons. This has been interpreted as 
the impossibility of using single-electron Fermi-Dirac distributions $f(\epsilon_i)$
in a transport theory devoted to correlated systems.\cite{DelaneyGreer:04a} 
This assertion might seem true, but it does not necessarily apply.
As is well known, electron Fermi distribution 
functions are ubiquitously employed in transport theories, 
ranging from the (semi)classical Boltzmann equation to the Keldysh NEGF formalism.
Whether for macroscopic, mesoscopic or nanoscopic systems, whether applied to uncorrelated 
or strongly correlated systems,
these theories have in common the assumptions that: 
\par
(a) a separation in device and electrodes is possible, implying 
sufficiently small device-electrode couplings, which insignificantly perturb 
the electrodes. The properties of the electrodes connected to the device 
do not differ from those of the isolated electrodes;
\par
(b) electron correlations in electrodes are negligible, which implies 
that, there, the electron distributions are Fermi functions. 
\par
Notice that only the electron distributions \emph{in electrodes} are constrained, 
more precisely, they are constrained to be the same as in the \emph{isolated} electrodes,
and there they are 
Fermi functions \emph{irrespective} whether the device 
is correlated or not. The Fermi distribution is the correct 
boundary condition for the uncorrrelated case. Because, as it will be shown below, 
with these boundary conditions the DG approach yields unphysical results, this suffices 
to demonstrate that the variational approach itself is incorrect. 
We do not intend to rigorously prove here that the Fermi distribution is the correct boundary condition 
for correlated nanoscopic/molecular systems. 
Still, we note that 
with boundary conditions expressed by Fermi functions one can successfully 
describe the electric transport in macroscopic systems. Moreover, by employing 
the same boundary conditions it was possible to explain 
delicate aspects of the transport in correlated nanosystems (e.~g., the unitary limit 
for the Kondo plateau  
in single-electron transistors) in good agreement with experiment.
Therefore, 
we argue that the description in terms of Fermi functions of the electrons in 
electrodes is plausible even if the latter are coupled to correlated devices.
\par
Let us now express the boundary conditions (\ref{eq-Q}) just 
in the aforementioned manner, i.~e., by using the Fermi distribution. 
In our case, condition (a) 
requires sufficiently weak couplings $t_{q_{L},\sigma}$  and $t_{q_{R},\sigma}$ 
between device and electrodes. In the absence of a voltage bias,
the reservoirs are in equilibrium among themselves ($\mu_L = \mu_R = \mu$). 
An applied bias $V$ does not affect the fact that each reservoir remains in local equilibrium,
but it drives them out of equilibrium with respect to one another,
because it shifts the chemical potentials, $\mu_L = \mu + V/2$ and $\mu_R = \mu - V/2$.
It is just the imbalance of the corresponding chemical potentials that 
is kept constant by an external power supply, which causes a steady-state current through the device.
\par
\begin{figure}[htb]
$ $\\[7ex]
\centerline{\hspace*{-0ex}
\includegraphics[width=0.70\textwidth,angle=-90]{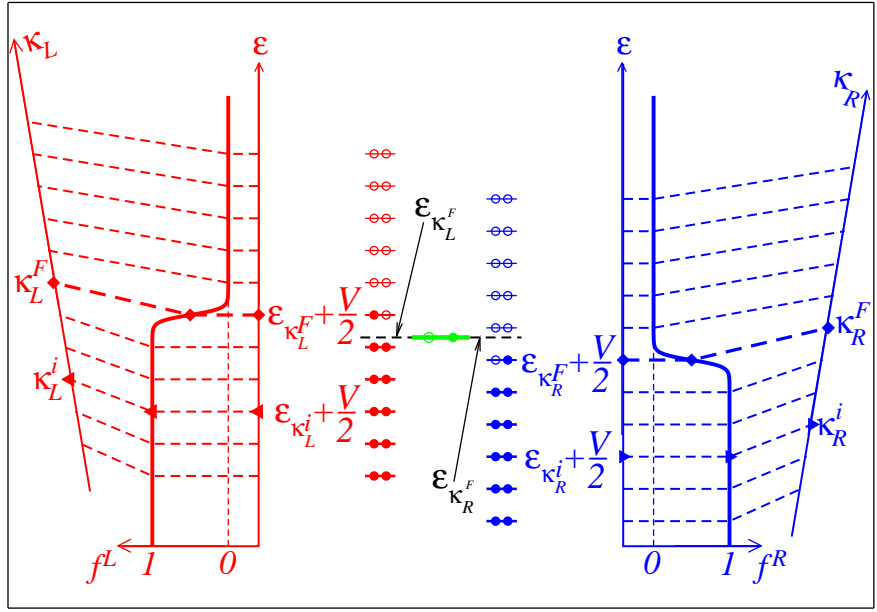}}
$ $\\[3ex]
\caption{(Color online) Effect of an applied bias $V$ on the single-electron 
distributions $f^L$ and $f^R$ in the left ($L$) and right ($R$) electrodes. 
At $V=0$, the electrodes are in equilibrium, and their chemical potentials are equal,
$\varepsilon_{{\kappa}_{L}^{F}} = \varepsilon_{{\kappa}_{R}^{F}} = \mu $ (black dashed horizontal line).
The bias $V$ creates an imbalance between the chemical potentials 
$\mu_L = \varepsilon_{{\kappa}_{L}^{F}} + V/2$ and 
$\mu_R = \varepsilon_{{\kappa}_{R}^{F}} - V/2 $
responsible for the current flow. The figure visualizes 
the fact that the bias only shifts 
the single-electron energies $\varepsilon_{{\kappa}_{L,R}}$ 
in the two electrodes by $\pm V/2$ in opposite directions: 
$\varepsilon_{{\kappa}_{L}^{i}}(V) = \varepsilon_{{\kappa}_{L,R}^{i}} + V/2$ and 
$\varepsilon_{{\kappa}_{R}^{i}}(V) = \varepsilon_{{\kappa}_{R}^{i}} - V/2$. 
However, the bias affects neither the labels of the single-electron states 
[$\kappa_{L,R}^{i}(V) \equiv \kappa_{L,R}^{i}$]
nor their occupancies: $f^{L,R}_{\kappa_{L,R}^{i}}(V) 
= f^{L,R}(\varepsilon_{{\kappa}_{L,R}^{i}} \pm V/2; V) 
= f^{L,R}(\varepsilon_{{\kappa}_{L,R}^{i}}; V=0) = f^{L,R}_{\kappa_{L,R}^{i}}(V=0) $ 
[\emph{cf.~}Eq.~(\ref{eq-nu})]. The representation in the figure 
is schematic: in reality, neither 
the energies $\varepsilon$ (vertical axes) nor the quantum indices 
$\kappa_{L,R}$ (oblique axes) have equidistant values.}
\label{fig}
\end{figure}

Let us assume that for $V=0$ the single-electron states in the left electrode, which are specified 
by the labels $\kappa_1, \kappa_2, \ldots $, have the energies 
$\varepsilon_{\kappa_1}, \varepsilon_{\kappa_2}, \ldots $.
(Obviously, the considerations presented below also apply to the right electrode.)
They can be obtained by diagonalizing the term $H_L$ of Eq.\ (\ref{eq-hamiltonian})
\begin{equation}
\label{eq-H_L}
H_L = \sum_{j} \varepsilon_{\kappa_j} \alpha_{\kappa_j}^{\dagger} \alpha_{\kappa_j} ,
\end{equation}
although for the present purposes we need not to specify 
the explicit transformation $\boldsymbol{\Gamma}$
\begin{equation}
\label{eq-Gamma}
a_{l\sigma} = \sum_{j} \Gamma_{\kappa_j,l \sigma} \alpha_{\kappa} .
\end{equation}
In the presence of a bias, 
the single-electron states remain specified by the same labels
$\kappa_1, \kappa_2, \ldots $: it is the same linear orthogonal transformation 
$\boldsymbol{\Gamma}$ that diagonalizes 
$H_L$ both for $\mu_L = \mu$ and for $\mu_L = \mu + V/2$.
The only change brought about by the bias 
is that the single-particle energies are simply shifted by $V/2$ with respect the former, 
$H_L = \sum_{j} (\varepsilon_{\kappa_j} + V/2)\alpha_{\kappa_j}^{\dagger} \alpha_{\kappa_j}$.
The occupancies of the single-particle states 
$f_{\kappa_j} = \langle \alpha_{\kappa_j}^{\dagger} \alpha_{\kappa_j} \rangle$ at equilibrium ($V=0$) 
and in the steady state ($V\neq 0$) are the same: $f^{L,R}_{\kappa_j}(V) = f^{L,R}_{0,\kappa_j} \equiv f^{L,R}_{\kappa_j}(V=0)$. 
Only their energy distributions are shifted  
$f^{L,R}(\varepsilon_{\kappa_j}) = f_0^{L,R}(\varepsilon_{\kappa_j} \mp V/2)$. 
The situation is schematically depicted in Fig.~\ref{fig}. Of course, the above considerations 
are not restricted to the linear response limit.
\par
By choosing now the electron occupancies to express the boundary conditions
($Q_{\kappa} \equiv \alpha_{\kappa}^{\dagger} \alpha_{\kappa}$) and 
assuming, as usual, that the electron distributions in the reservoirs are not 
affected by the coupling to the device, we can reexpress Eqs.\ (\ref{eq-Q}) as
\begin{equation}  
\label{eq-nu}  
\left \langle \Psi \right \vert \alpha_{\kappa}^{\dagger} \alpha_{\kappa} \left \vert \Psi \right \rangle 
= 
\left \langle \Psi_{0} \right \vert \alpha_{\kappa}^{\dagger} \alpha_{\kappa} \left \vert \Psi_0 \right \rangle ,
\end{equation}  
which apply both to the left and the right electrodes. 
Most importantly, the quantities 
$Q_{\kappa, n} = \mathcal{X}_{\kappa, n} = 
\left\langle \Psi_{n} \right \vert  \alpha_{\kappa}^{\dagger} \alpha_{\kappa} \left \vert \Psi_{0} \right \rangle $
entering the l.h.s.\ of Eq.\ (\ref{eq-Q-n}) are real. With this choice, 
$\mathcal{Y}_{\kappa, n} \equiv 0$, and  
Eqs.\ (\ref{eq-lambda}) and (\ref{eq-chi}) become two sets of decoupled equations
for $\lambda$'s and $\chi$'s, respectively
\begin{eqnarray}
\displaystyle
\label{eq-lambda-nu}  
& & \sum_{\kappa^\prime} \lambda_{\kappa^{\prime}} 
\sum_{n\neq 0} 
\frac{\mathcal{X}_{\kappa^{\prime},n} \mathcal{X}_{\kappa,n} }{E_{n} - E_{0}}
=
\sum_{n\neq 0} \frac{\mathcal{W}_{n} }{E_{n} - E_{0}} \mathcal{J}_{q,n} , \\
\label{eq-chi-nu}  
& & 
\sum_{q^\prime} \chi_{q^\prime} \sum_{n\neq 0} \frac{\mathcal{J}_{q^\prime, n} \mathcal{J}_{q, n} }{E_{n} - E_{0}}
= 0 .
\end{eqnarray}  
From the latter equations we get $\chi_{q} = 0$,\cite{nonvanishing-determinants} 
and then Eq.\ (\ref{eq-J}) yields
\begin{equation}
\label{eq-J=0}
J = 0 .
\end{equation}
This means that the approach considered above predicts a vanishing 
zero-bias conductance ($G=J/V=0$) at least for the rather general models of the type 
(\ref{eq-hamiltonian}). One should note that this prediction also holds if 
not all $\kappa$'s, but only some of the occupancies of the single-electron states 
are subject to constraints of the type (\ref{eq-nu}). 
Furthermore, the prediction (\ref{eq-J=0}) also applies to the case where 
the constraints (\ref{eq-j-constraints}) are not applied 
for \emph{all} the sites $q$, but, e.~g., only at the two contacts (i.~e., 
the current flowing from the left electrode into the device is equal 
to that from the device into the right electrode).
\par
Needless to say, the prediction (\ref{eq-J=0}) of a vanishing current 
in the linear response limit (vanishing zero-bias conductance) 
is completely unphysical and flagrantly contradicts 
numerous well established results.  
The existence of a nonvanishing zero-bias conductance $G = G_0 \equiv 2e^2/h$ 
in the textbook example of conduction through a resonant single level 
system,\cite{Datta:05} or the fact that the same value $G_0$ characterizes the 
Kondo plateau in the transport through a single-electron transistor 
\cite{Izumida:01,Costi:01} are only two particular cases, which 
are described by the general model (\ref{eq-hamiltonian}) 
for which the above considerations apply.
\par
Concerning the linear response limit of the variational approach discussed 
here, we still have the following comments. Without imposing any 
boundary conditions of the type (\ref{eq-Q}), the minimization of 
$\mathcal{E}$ of Eq.\ (\ref{eq-total-energy}) yields equations identical 
to Eqs.\ (\ref{eq-A}) and (\ref{eq-chi}) if we set $\lambda_{\kappa} \equiv 0$.
Because in this case the latter group of equations only possess the trivial solution 
$\chi_{q} = 0$, the expansion coefficients obtained from Eq.\ (\ref{eq-A})
have the form $A_{n} =  \mathcal{A}_{n} \equiv \mathcal{W}_{n}/(E_0 - E_n) $, 
\emph{i.~e.}, the solution 
$\vert \Psi \rangle$ of Eq.\ (\ref{eq-Psi}) is nothing but the ground state 
$\vert \mathcal{G}\rangle$ of the total Hamiltonian $H + W$ obtained 
within the first-order perturbation theory with respect to $W$. 
Being the ground state, 
that is, an eigenstate of the total Hamiltonian, 
it obviously obeys the equation of continuity: the current is site independent, 
more precisely $J=J_q=\langle \mathcal{G}\vert j_{q}\vert\mathcal{G}\rangle = 0$.\cite{beyond-LRA}
When the supplementary (boundary) conditions (\ref{eq-Q}) are imposed, the corresponding 
Lagrange multipliers become nonvanishing $\lambda_{\kappa} \neq 0$, as 
seen from Eq.\ (\ref{eq-lambda}) or even from the more particular Eq.\ (\ref{eq-lambda-nu}).
Consequently, $A_{n} \neq \mathcal{A}_{n}$, that is, the solution of optimization 
does differ from the ground state with applied bias, 
$\vert \Psi\rangle \neq \vert \mathcal{G}\rangle$. 
The essential point is that, as we have seen above, 
in order to sustain a nonvanishing current ($J\neq 0$), the 
matrix elements of the operators employed to impose the boundary conditions 
should have nonvanishing imaginary parts, 
$\mbox{Im\,}Q_{\kappa,n} \equiv \mathcal{Y}_{\kappa,n} \neq 0$.
Luckily, this happens to be the case if the Fano operator (\ref{eq-fano}) is used 
in Eq.\ (\ref{eq-Q}), which amounts 
to formulate the boundary conditions in terms of the Wigner function.\cite{Baldea:2008b}
This choice yields nonvanishing currents, which sometimes, by chance, could be comparable 
with measured values \cite{DelaneyGreer:04a,DelaneyGreer:04b}
or look plausible.\cite{Fagas:06,Fagas:07} 
However, as we have recently unambiguously demonstrated,\cite{Baldea:2008b} 
the variational approach based on the Wigner function 
fails to recover the simplest results both for uncorrelated and correlated nanosystems,
where its predictions are completely unphysical.
\section{Discussion and conclusion}
\label{sec-conclusion}
In Ref.~\onlinecite{Baldea:2008b}, we demonstrated that 
the DG approach \cite{DelaneyGreer:04a,DelaneyGreer:04b,DelaneyGreer:06}
fails to reproduce the simplest well established results for transport in nanosystems
and especially criticized the boundary conditions.
In view of the fact that a reliable method to deal with correlated electron transport 
is hardly needed, one might still think to be able to develop the valid approach  
based on the DG method but employing other boundary conditions.
Therefore, in the present paper, we have inquired the validity of the DG variational approach 
by considering constraints other than those imposed to
the Wigner function at the boundaries. 
We believe that this investigation is important for specialists in the field thinking to apply 
such a modified DG scheme: the implementation for realistic 
ab initio calculations is by no means an easy task, and this probably explains why
no other group applied the DG method in spite of its claimed success.
\par
In the first part we have examined the linear response approximation 
by considering general boundary conditions. Further, we have critically analyzed 
the claim of DG on the Fermi-Dirac single-electron distributions. 
Namely, they claimed \cite{DelaneyGreer:04a,DelaneyGreer:04b,DelaneyGreer:06} 
that electron Fermi distribution 
functions, which are ubiquitously employed in transport theories, cannot be 
used in correlated systems because
the single-particle picture breaks down, and resorting to the Wigner function 
was proposed as a way out of this difficulty.
In Sect.\  \ref{sec-Fermi-BC}, we have explained that strong 
electron correlations \emph{in devices} do not preclude the usage of single-electron distributions 
to express boundary conditions. 
The reason is that they should be imposed 
\emph{in electrodes}, and there electrons are (assumed to be) uncorrelated.
In fact, constraining single-electron distributions to be of Fermi-Dirac 
type in electrodes is the common feature of the most transport theories employed for
macroscopic, mesoscopic and nanoscopic scale, and of interest there is just 
the case of correlated systems. 
\par
Obvioulsy, it is possible to impose constraints on the single-electron distributions
in uncorrelated systems. As an important point of our analysis, we have demonstrated that, in particular, 
even for such uncorrelated systems and these standard boundary conditions
(a case beyond of all question) the variational approach predicts a 
vanishing zero-bias conductance, which represents a completely unphysical result. 
Although this very fact is sufficient to invalidate this variational approach,  
the conclusion of the present investigation applies to more much general situations where
strong electron correlations are present.
Such more general examples include (but are not limited to) the systems 
described by the model Hamiltonian (\ref{eq-hamiltonian}).
\par
In principle, any theoretical approach of transport is based on two main ingredients. 
(i) First, one has to determine a certain 
quantity, which characterizes the transport cluster. 
(ii) Second, one has to impose boundary conditions to link the cluster to 
electrodes. The first ingredient is e.~g., the transmission coefficient,
Green function(s), or transition probabilities in Ladauer, NEGF, or master 
equation approaches, respectively. 
While this first ingredient is different from one transport approach to another,
the second is common for virtually all approaches: whether correlated or not, the cluster is 
linked to electrodes by employing Fermi distributions. 
Because none of these approaches yields 
results which are manifestly unphysical, one has to admit that the formulation of the 
boundary conditions in terms of Fermi distribution functions is legitimate, or at least does not 
lead to manifest absurdities. If these boundary conditions were wrong, all these 
widely employed 
approaches would also yield unphysical results, but this is not the case.
In fact, it is hard to imagine a physical situation more typical for using
single-particle Fermi distributions than to describe electrons in metals (electrodes).
\par
The approach of Delaney and Greer also comprises these two aspects: 
(i) One minimizes the total energy $E$ of the finite cluster used for transport calculations 
in the presence of an applied voltage. The energy of this open system 
is computed as the average value of a Hermitian Hamiltonian, $E = \langle H + W\rangle$.
(ii) This is a constrained minimization: specifically, boundary conditions are 
applied to the Wigner function.
\par
In Ref.~\onlinecite{Baldea:2008b}, we demonstrated that, 
in the form proposed in Ref.~\onlinecite{DelaneyGreer:04a},
this approach yields unphysical results and 
especially criticized the boundary conditions 
adopted by Delaney and Greer in the context of \emph{their} approach.
In the present paper, we have lifted these specific constraints and replaced 
them with the widely used boundary conditions, expressed in terms of Fermi functions.
The result obtained, a vanishingly zero-bias conductance, is again quite unphysical.
So, even with these usual boundary conditions, formulated in terms of 
Fermi distributions in electrodes, in the way common to the most widely employed
transport theories, the approach 
based on the DG-variational method fails to pass the minimal 
decisive test, which any approach of electric transport at nanoscale must satisfy;
namely, to be at least able to
correctly describe the conductance in the linear response (Kubo) approximation 
or even to reproduce well established results of the Landauer theory in the
absence of correlations (see, for example, Ref.\ \onlinecite{BaerNeuhauser:03}).
\par
The main result of our study is that the DG-variational approach does not work
even with modified boundary conditions. This implies that the DG-approach 
is invalid not (or not only) because of the boundary conditions.
If the imposition of the boundary conditions 
in terms of Wigner function were the only wrong
point, the variational approach would 
produce correct results 
with correct boundary conditions. Doubtless, the Fermi functions represent 
the correct boundaries for uncorrelated system, a fact which proves the  
failure of the DG approach for that case. If the DG variational approach is incorrect even for uncorrelated systems, 
it is inconceivable that it works for correlated systems.
We are then necessarily led to the conclusion that the other ingredient (i) of the 
DG-variational approach, namely the minimization of the cluster's total
energy, computed as the average of a Hermitian Hamiltonian, is inappropriate.
\par
So, the failure of the approach of Delaney and Greer 
cannot be solely assigned to the manner in which the boundary conditions 
are imposed. Constraining the Wigner function $f(q,p)$ at the boundaries 
can also produce valuable physical results for the steady state current, 
provided that $f(q,p)$ is determined from the Liouville equation for 
$\partial f/\partial t = 0$ and the values $f(q_{L,R}, p)$ at the boundaries determined by the 
Fermi functions in reservoirs. \cite{frensley:91}
However, in that case, the fact that the investigated systems are open 
leads to a non-Hermitian Liouville operator, which yields time irreversibility 
and currents that saturate to values characterizing the steady state flow.
The counterpart of this procedure would be to employ non-Hermitian Hamilton operators 
$\mathcal{H} \neq \mathcal{H}^\dagger$,
amounting to consider imaginary self-energies resulting from electrode-device interactions. 
\cite{Datta:05} Attempting to develop a variational approach
for a finite open system (device) described by a non-Hermitian Hamiltonian, 
e.~g., by minimizing 
$\left\langle \left(\mathcal{H} + W - E \right) \left(\mathcal{H}^\dagger + W - E \right) \right\rangle$
instead of Eq.\ (\ref{eq-total-energy}), might be an 
interesting alternative to 
the existing approaches to correlated transport, but to our knowledge 
such a method is not available at present. 
Till then, one is forced to admit that the variational procedure 
attempting to obtain the steady state current through 
a finite open system described by a Hermitian Hamiltonian 
and a wave function that minimizes 
the total energy in the presence of a voltage bias with certain boundary conditions 
is unable to reproduce
well established results, on which there is an incontestable agreement 
between experiments and other theoretical approaches. 
\section*{Acknowledgments}
The authors acknowledge with thanks the financial support for this work 
provided by the Deu\-tsche For\-schungs\-ge\-mein\-schaft (DFG).

\begin{thebibliography}{38}
\expandafter\ifx\csname natexlab\endcsname\relax\def\natexlab#1{#1}\fi
\expandafter\ifx\csname bibnamefont\endcsname\relax
  \def\bibnamefont#1{#1}\fi
\expandafter\ifx\csname bibfnamefont\endcsname\relax
  \def\bibfnamefont#1{#1}\fi
\expandafter\ifx\csname citenamefont\endcsname\relax
  \def\citenamefont#1{#1}\fi
\expandafter\ifx\csname url\endcsname\relax
  \def\url#1{\texttt{#1}}\fi
\expandafter\ifx\csname urlprefix\endcsname\relax\def\urlprefix{URL }\fi
\providecommand{\bibinfo}[2]{#2}
\providecommand{\eprint}[2][]{\url{#2}}

\bibitem[{\citenamefont{Heurich et~al.}(2002)\citenamefont{Heurich, Cuevas,
  Wenzel, and Sch\"on}}]{Wenzel:02}
\bibinfo{author}{\bibfnamefont{J.}~\bibnamefont{Heurich}},
  \bibinfo{author}{\bibfnamefont{J.~C.} \bibnamefont{Cuevas}},
  \bibinfo{author}{\bibfnamefont{W.}~\bibnamefont{Wenzel}}, \bibnamefont{and}
  \bibinfo{author}{\bibfnamefont{G.}~\bibnamefont{Sch\"on}},
  \bibinfo{journal}{Phys. Rev. Lett.} \textbf{\bibinfo{volume}{88}},
  \bibinfo{pages}{256803} (\bibinfo{year}{2002}).

\bibitem[{\citenamefont{Nitzan and Ratner}(2003)}]{Nitzan:03}
\bibinfo{author}{\bibfnamefont{A.}~\bibnamefont{Nitzan}} \bibnamefont{and}
  \bibinfo{author}{\bibfnamefont{M.~A.} \bibnamefont{Ratner}},
  \bibinfo{journal}{Science} \textbf{\bibinfo{volume}{300}},
  \bibinfo{pages}{1384} (\bibinfo{year}{2003}).

\bibitem[{\citenamefont{Stokbro et~al.}(2003)\citenamefont{Stokbro, Taylor,
  Brandbyge, Mozos, and Ordejón}}]{Stokbro:03}
\bibinfo{author}{\bibfnamefont{K.}~\bibnamefont{Stokbro}},
  \bibinfo{author}{\bibfnamefont{J.}~\bibnamefont{Taylor}},
  \bibinfo{author}{\bibfnamefont{M.}~\bibnamefont{Brandbyge}},
  \bibinfo{author}{\bibfnamefont{J.-L.} \bibnamefont{Mozos}}, \bibnamefont{and}
  \bibinfo{author}{\bibfnamefont{P.}~\bibnamefont{Ordejón}},
  \bibinfo{journal}{Comp. Mat. Sci.} \textbf{\bibinfo{volume}{27}},
  \bibinfo{pages}{151} (\bibinfo{year}{2003}).

\bibitem[{\citenamefont{Baksmaty et~al.}(2008)\citenamefont{Baksmaty,
  Yannouleas, and Landman}}]{baksmaty:136803}
\bibinfo{author}{\bibfnamefont{L.~O.} \bibnamefont{Baksmaty}},
  \bibinfo{author}{\bibfnamefont{C.}~\bibnamefont{Yannouleas}},
  \bibnamefont{and} \bibinfo{author}{\bibfnamefont{U.}~\bibnamefont{Landman}},
  \bibinfo{journal}{Phys. Rev. Lett.} \textbf{\bibinfo{volume}{101}},
  \bibinfo{eid}{136803} (\bibinfo{year}{2008}).

\bibitem[{NPJ()}]{NPJ-9-2007}
\bibinfo{note}{See, for example, New J. Phys {\bf 9} (2007), especially the focus articles 111-125}.

\bibitem[{\citenamefont{Landauer}(1970)}]{Landauer:70}
\bibinfo{author}{\bibfnamefont{R.}~\bibnamefont{Landauer}},
  \bibinfo{journal}{Phil. Mag.} \textbf{\bibinfo{volume}{21}},
  \bibinfo{pages}{863 } (\bibinfo{year}{1970}).

\bibitem[{\citenamefont{B\"uttiker et~al.}(1985)\citenamefont{B\"uttiker, Imry,
  Landauer, and Pinhas}}]{Buttiker:85}
\bibinfo{author}{\bibfnamefont{M.}~\bibnamefont{B\"uttiker}},
  \bibinfo{author}{\bibfnamefont{Y.}~\bibnamefont{Imry}},
  \bibinfo{author}{\bibfnamefont{R.}~\bibnamefont{Landauer}}, \bibnamefont{and}
  \bibinfo{author}{\bibfnamefont{S.}~\bibnamefont{Pinhas}},
  \bibinfo{journal}{Phys. Rev. B} \textbf{\bibinfo{volume}{31}},
  \bibinfo{pages}{6207} (\bibinfo{year}{1985}).

\bibitem[{\citenamefont{Haug and Jauho}(1996)}]{HaugJauho}
\bibinfo{author}{\bibfnamefont{H.}~\bibnamefont{Haug}} \bibnamefont{and}
  \bibinfo{author}{\bibfnamefont{A.-P.} \bibnamefont{Jauho}},
  \emph{\bibinfo{title}{Quantum Kinetics in Transport and Optics of
  Semiconductors}}, vol. \bibinfo{volume}{123} (\bibinfo{publisher}{Springer
  Series in Solid-State Sciences}, \bibinfo{address}{Berlin, Heidelberg, New
  York}, \bibinfo{year}{1996}).

\bibitem[{\citenamefont{Datta}(2005)}]{Datta:05}
\bibinfo{author}{\bibfnamefont{S.}~\bibnamefont{Datta}},
  \emph{\bibinfo{title}{Quantum Transport: Atom to Transistor}}
  (\bibinfo{publisher}{Cambridge Univ. Press}, \bibinfo{year}{2005}).


\bibitem[{\citenamefont{White and Feiguin}(2004)}]{White:04}
\bibinfo{author}{\bibfnamefont{S.~R.} \bibnamefont{White}} \bibnamefont{and}
  \bibinfo{author}{\bibfnamefont{A.~E.} \bibnamefont{Feiguin}},
  \bibinfo{journal}{Phys. Rev. Lett.} \textbf{\bibinfo{volume}{93}},
  \bibinfo{pages}{076401} (\bibinfo{year}{2004}).

\bibitem[{\citenamefont{Daley et~al.}(2004)\citenamefont{Daley, Kollath,
  Schollw\"{o}ck, and Vidal}}]{Daley:04}
\bibinfo{author}{\bibfnamefont{A.~J.} \bibnamefont{Daley}},
  \bibinfo{author}{\bibfnamefont{C.}~\bibnamefont{Kollath}},
  \bibinfo{author}{\bibfnamefont{U.}~\bibnamefont{Schollw\"{o}ck}},
  \bibnamefont{and} \bibinfo{author}{\bibfnamefont{G.}~\bibnamefont{Vidal}},
  \bibinfo{journal}{Journal of Statistical Mechanics: Theory and Experiment}
  \textbf{\bibinfo{volume}{2004}}, \bibinfo{pages}{P04005}
  (\bibinfo{year}{2004}).

\bibitem[{\citenamefont{Feiguin and White}(2005)}]{White:05}
\bibinfo{author}{\bibfnamefont{A.~E.} \bibnamefont{Feiguin}} \bibnamefont{and}
  \bibinfo{author}{\bibfnamefont{S.~R.} \bibnamefont{White}},
  \bibinfo{journal}{Phys. Rev. B} \textbf{\bibinfo{volume}{72}},
  \bibinfo{eid}{020404} (\bibinfo{year}{2005}).

\bibitem[{\citenamefont{Bulla et~al.}(2008)\citenamefont{Bulla, Costi, and
  Pruschke}}]{Bulla:08}
\bibinfo{author}{\bibfnamefont{R.}~\bibnamefont{Bulla}},
  \bibinfo{author}{\bibfnamefont{T.~A.} \bibnamefont{Costi}}, \bibnamefont{and}
  \bibinfo{author}{\bibfnamefont{T.}~\bibnamefont{Pruschke}},
  \bibinfo{journal}{Rev. Mod. Phys.} \textbf{\bibinfo{volume}{80}},
  \bibinfo{eid}{395} (\bibinfo{year}{2008}).

\bibitem[{\citenamefont{Taylor et~al.}(2001)\citenamefont{Taylor, Guo, and
  Wang}}]{Taylor:01}
\bibinfo{author}{\bibfnamefont{J.}~\bibnamefont{Taylor}},
  \bibinfo{author}{\bibfnamefont{H.}~\bibnamefont{Guo}}, \bibnamefont{and}
  \bibinfo{author}{\bibfnamefont{J.}~\bibnamefont{Wang}},
  \bibinfo{journal}{Phys. Rev. B} \textbf{\bibinfo{volume}{63}},
  \bibinfo{pages}{245407} (\bibinfo{year}{2001}).

\bibitem[{\citenamefont{Brandbyge et~al.}(2002)\citenamefont{Brandbyge, Mozos,
  Ordej\'on, Taylor, and Stokbro}}]{Brandbyge:02}
\bibinfo{author}{\bibfnamefont{M.}~\bibnamefont{Brandbyge}},
  \bibinfo{author}{\bibfnamefont{J.-L.} \bibnamefont{Mozos}},
  \bibinfo{author}{\bibfnamefont{P.}~\bibnamefont{Ordej\'on}},
  \bibinfo{author}{\bibfnamefont{J.}~\bibnamefont{Taylor}}, \bibnamefont{and}
  \bibinfo{author}{\bibfnamefont{K.}~\bibnamefont{Stokbro}},
  \bibinfo{journal}{Phys. Rev. B} \textbf{\bibinfo{volume}{65}},
  \bibinfo{pages}{165401} (\bibinfo{year}{2002}).

\bibitem[{\citenamefont{Rocha et~al.}(2006)\citenamefont{Rocha,
  Garc\'{\i}a-Su\'{a}rez, Bailey, Lambert, Ferrer, and Sanvito}}]{Rocha:06}
\bibinfo{author}{\bibfnamefont{A.~R.} \bibnamefont{Rocha}},
  \bibinfo{author}{\bibfnamefont{V.~M.} \bibnamefont{Garc\'{\i}a-Su\'{a}rez}},
  \bibinfo{author}{\bibfnamefont{S.}~\bibnamefont{Bailey}},
  \bibinfo{author}{\bibfnamefont{C.}~\bibnamefont{Lambert}},
  \bibinfo{author}{\bibfnamefont{J.}~\bibnamefont{Ferrer}}, \bibnamefont{and}
  \bibinfo{author}{\bibfnamefont{S.}~\bibnamefont{Sanvito}},
  \bibinfo{journal}{Phys. Rev. B} \textbf{\bibinfo{volume}{73}},
  \bibinfo{eid}{085414} (\bibinfo{year}{2006}).

\bibitem[{\citenamefont{Hettler et~al.}(2003)\citenamefont{Hettler, Wenzel,
  Wegewijs, and Schoeller}}]{Wenzel:03}
\bibinfo{author}{\bibfnamefont{M.~H.} \bibnamefont{Hettler}},
  \bibinfo{author}{\bibfnamefont{W.}~\bibnamefont{Wenzel}},
  \bibinfo{author}{\bibfnamefont{M.~R.} \bibnamefont{Wegewijs}},
  \bibnamefont{and}
  \bibinfo{author}{\bibfnamefont{H.}~\bibnamefont{Schoeller}},
  \bibinfo{journal}{Phys. Rev. Lett.} \textbf{\bibinfo{volume}{90}},
  \bibinfo{pages}{076805} (\bibinfo{year}{2003}).

\bibitem[{\citenamefont{Muralidharan et~al.}(2006)\citenamefont{Muralidharan,
  Ghosh, and Datta}}]{Muralidharan:06}
\bibinfo{author}{\bibfnamefont{B.}~\bibnamefont{Muralidharan}},
  \bibinfo{author}{\bibfnamefont{A.~W.} \bibnamefont{Ghosh}}, \bibnamefont{and}
  \bibinfo{author}{\bibfnamefont{S.}~\bibnamefont{Datta}},
  \bibinfo{journal}{Phys. Rev. B} \textbf{\bibinfo{volume}{73}},
  \bibinfo{eid}{155410} (\bibinfo{year}{2006}).

\bibitem[{\citenamefont{Thygesen and Rubio}(2007)}]{Thygesen:07}
\bibinfo{author}{\bibfnamefont{K.~S.} \bibnamefont{Thygesen}} \bibnamefont{and}
  \bibinfo{author}{\bibfnamefont{A.}~\bibnamefont{Rubio}}, \bibinfo{journal}{J.
  Chem. Phys.} \textbf{\bibinfo{volume}{126}}, \bibinfo{eid}{091101}
  (\bibinfo{year}{2007}).

\bibitem[{\citenamefont{Reed et~al.}(1997)\citenamefont{Reed, Zhou, Muller,
  Burgin, and Tour}}]{Reed:97}
\bibinfo{author}{\bibfnamefont{M.~A.} \bibnamefont{Reed}},
  \bibinfo{author}{\bibfnamefont{C.}~\bibnamefont{Zhou}},
  \bibinfo{author}{\bibfnamefont{C.~J.} \bibnamefont{Muller}},
  \bibinfo{author}{\bibfnamefont{T.~P.} \bibnamefont{Burgin}},
  \bibnamefont{and} \bibinfo{author}{\bibfnamefont{J.~M.} \bibnamefont{Tour}},
  \bibinfo{journal}{Science} \textbf{\bibinfo{volume}{278}},
  \bibinfo{pages}{252} (\bibinfo{year}{1997}).

\bibitem[{\citenamefont{Xiao et~al.}(2004)\citenamefont{Xiao, Xu, and
  Tao}}]{Xiao:04}
\bibinfo{author}{\bibfnamefont{X.}~\bibnamefont{Xiao}},
  \bibinfo{author}{\bibfnamefont{B.}~\bibnamefont{Xu}}, \bibnamefont{and}
  \bibinfo{author}{\bibfnamefont{N.}~\bibnamefont{Tao}}, \bibinfo{journal}{Nano
  Letters} \textbf{\bibinfo{volume}{4}}, \bibinfo{pages}{267}
  (\bibinfo{year}{2004}).

\bibitem[{\citenamefont{Dadosh et~al.}(2005)\citenamefont{Dadosh, Gordin,
  Krahne, Khivrich, Mahalu, Frydman, Sperling, Yacoby, and
  Bar-Joseph}}]{Dadosh:05}
\bibinfo{author}{\bibfnamefont{T.}~\bibnamefont{Dadosh}},
  \bibinfo{author}{\bibfnamefont{Y.}~\bibnamefont{Gordin}},
  \bibinfo{author}{\bibfnamefont{R.}~\bibnamefont{Krahne}},
  \bibinfo{author}{\bibfnamefont{I.}~\bibnamefont{Khivrich}},
  \bibinfo{author}{\bibfnamefont{D.}~\bibnamefont{Mahalu}},
  \bibinfo{author}{\bibfnamefont{V.}~\bibnamefont{Frydman}},
  \bibinfo{author}{\bibfnamefont{J.}~\bibnamefont{Sperling}},
  \bibinfo{author}{\bibfnamefont{A.}~\bibnamefont{Yacoby}}, \bibnamefont{and}
  \bibinfo{author}{\bibfnamefont{I.}~\bibnamefont{Bar-Joseph}},
  \bibinfo{journal}{Nature} \textbf{\bibinfo{volume}{436}}, \bibinfo{pages}{677
  } (\bibinfo{year}{2005}).

\bibitem[{\citenamefont{Tsutsui et~al.}(2006)\citenamefont{Tsutsui, Teramae,
  Kurokawa, and Sakai}}]{tsutsui:06}
\bibinfo{author}{\bibfnamefont{M.}~\bibnamefont{Tsutsui}},
  \bibinfo{author}{\bibfnamefont{Y.}~\bibnamefont{Teramae}},
  \bibinfo{author}{\bibfnamefont{S.}~\bibnamefont{Kurokawa}}, \bibnamefont{and}
  \bibinfo{author}{\bibfnamefont{A.}~\bibnamefont{Sakai}},
  \bibinfo{journal}{Appl. Phys. Lett.} \textbf{\bibinfo{volume}{89}},
  \bibinfo{eid}{163111} (\bibinfo{year}{2006}).

\bibitem[{\citenamefont{Delaney and
  Greer}(2004{\natexlab{a}})}]{DelaneyGreer:04a}
\bibinfo{author}{\bibfnamefont{P.}~\bibnamefont{Delaney}} \bibnamefont{and}
  \bibinfo{author}{\bibfnamefont{J.~C.} \bibnamefont{Greer}},
  \bibinfo{journal}{Phys. Rev. Lett.} \textbf{\bibinfo{volume}{93}},
  \bibinfo{pages}{036805} (\bibinfo{year}{2004}{\natexlab{a}}).

\bibitem[{\citenamefont{Delaney and
  Greer}(2004{\natexlab{b}})}]{DelaneyGreer:04b}
\bibinfo{author}{\bibfnamefont{P.}~\bibnamefont{Delaney}} \bibnamefont{and}
  \bibinfo{author}{\bibfnamefont{J.~C.} \bibnamefont{Greer}},
  \bibinfo{journal}{Int. J. Quant. Chem.} \textbf{\bibinfo{volume}{100}},
  \bibinfo{pages}{1163} (\bibinfo{year}{2004}{\natexlab{b}}).

\bibitem[{\citenamefont{Delaney and Greer}(2006)}]{DelaneyGreer:06}
\bibinfo{author}{\bibfnamefont{P.}~\bibnamefont{Delaney}} \bibnamefont{and}
  \bibinfo{author}{\bibfnamefont{J.~C.} \bibnamefont{Greer}},
  \bibinfo{journal}{Proc. Roy. Soc. A} \textbf{\bibinfo{volume}{462}},
  \bibinfo{pages}{117} (\bibinfo{year}{2006}).

\bibitem[{\citenamefont{Fagas et~al.}(2006)\citenamefont{Fagas, Delaney, and
  Greer}}]{Fagas:06}
\bibinfo{author}{\bibfnamefont{G.}~\bibnamefont{Fagas}},
  \bibinfo{author}{\bibfnamefont{P.}~\bibnamefont{Delaney}}, \bibnamefont{and}
  \bibinfo{author}{\bibfnamefont{J.~C.} \bibnamefont{Greer}},
  \bibinfo{journal}{Phys. Rev. B} \textbf{\bibinfo{volume}{73}},
  \bibinfo{pages}{241314(R)} (\bibinfo{year}{2006}).

\bibitem[{\citenamefont{Fagas and Greer}(2007)}]{Fagas:07}
\bibinfo{author}{\bibfnamefont{G.}~\bibnamefont{Fagas}} \bibnamefont{and}
  \bibinfo{author}{\bibfnamefont{J.~C.} \bibnamefont{Greer}},
  \bibinfo{journal}{Nanotechnology} \textbf{\bibinfo{volume}{18}},
  \bibinfo{pages}{424010 (\bibinfo{year}{2007})}.

\bibitem[{\citenamefont{B\^aldea and K\"oppel}(2008)}]{Baldea:2008b}
\bibinfo{author}{\bibfnamefont{I.}~\bibnamefont{B\^aldea}} \bibnamefont{and}
  \bibinfo{author}{\bibfnamefont{H.}~\bibnamefont{K\"oppel}},
  \bibinfo{journal}{Phys. Rev. B} \textbf{\bibinfo{volume}{78}},
  \bibinfo{eid}{115315} (\bibinfo{year}{2008}).

\bibitem[{her()}]{hermitian}
\bibinfo{note}{Any operator $Q$ can be decomposed in a Hermitian and an
  anti-Hermitian part, $Q = \left(A + i B\right)/2 $ where $A = A^{\dagger}$ and
  $B = B^{\dagger}$ are Hermitian operators ($A \equiv Q + Q^{\dagger} $, $i B
  \equiv Q - Q^{\dagger}$). Because the complex conjugate of Eq.\ (\ref{eq-Q})
  must also be satisfied, $A$ and $B$ must satisfy similar conditions, $\langle
  \Psi \vert A \vert \Psi \rangle = \langle \Psi_{0} \vert A \vert \Psi_{0}
  \rangle$ and $\langle \Psi \vert B \vert \Psi \rangle = \langle \Psi_{0}
  \vert B \vert \Psi_{0} \rangle$.}

\bibitem[{\citenamefont{Caroli et~al.}(1971)\citenamefont{Caroli, Combescot,
  Nozi\`eres, and Saint-James}}]{Caroli:71}
\bibinfo{author}{\bibfnamefont{C.}~\bibnamefont{Caroli}},
  \bibinfo{author}{\bibfnamefont{R.}~\bibnamefont{Combescot}},
  \bibinfo{author}{\bibfnamefont{P.}~\bibnamefont{Nozi\`eres}},
  \bibnamefont{and}
  \bibinfo{author}{\bibfnamefont{D.}~\bibnamefont{Saint-James}},
  \bibinfo{journal}{J. Phys. C: Solid State Physics}
  \textbf{\bibinfo{volume}{4}}, \bibinfo{pages}{916} (\bibinfo{year}{1971}).

\bibitem[{non()}]{nonvanishing-determinants}
\bibinfo{note}{We checked by straightforward numerical calculations that these
  determinants do not vanish for the case of the point contact attached to long
  electrodes and for the single electron transistor connected to short
  electrodes.}

\bibitem[{gri()}]{grid}
\bibinfo{note}{In ab initio calculations like those of
  Refs.~\onlinecite{DelaneyGreer:04a,DelaneyGreer:04b,DelaneyGreer:06}, the
  equation of continuity (\ref{eq-j-constraints}) has to be imposed not only
  for a cluster of certain size ($N$), but also on a spatial grid containing a
  finite number of points $\{x_{l}\}_{l=1}^{N_g}$, and $N_g$ should also be
  sufficiently large to ensure convergence.}

\bibitem[{siz()}]{finite-N}
\bibinfo{note}{For example, $\mathcal{N} = {{2N}\choose{N}}$ in a half-filled cluster,
where the number of electrons is equal to the number of sites.}

\bibitem[{\citenamefont{Izumida et~al.}(2001)\citenamefont{Izumida, Sakai, and
  Suzuki}}]{Izumida:01}
\bibinfo{author}{\bibfnamefont{W.}~\bibnamefont{Izumida}},
  \bibinfo{author}{\bibfnamefont{O.}~\bibnamefont{Sakai}}, \bibnamefont{and}
  \bibinfo{author}{\bibfnamefont{S.}~\bibnamefont{Suzuki}},
  \bibinfo{journal}{J. Phys. Soc. Jpn.} \textbf{\bibinfo{volume}{70}},
  \bibinfo{pages}{1045} (\bibinfo{year}{2001}).

\bibitem[{\citenamefont{Costi}(2001)}]{Costi:01}
\bibinfo{author}{\bibfnamefont{T.~A.} \bibnamefont{Costi}},
  \bibinfo{journal}{Phys. Rev. B} \textbf{\bibinfo{volume}{64}},
  \bibinfo{pages}{241310(R)} (\bibinfo{year}{2001}).

\bibitem[{bey()}]{beyond-LRA}
\bibinfo{note}{In fact, it is easy to understand that this result is not
  limited to the linear response approximation: it is just the exact ground
  state $\vert \mathcal{G}\rangle$ of $H + W$, which minimizes $\mathcal{E}$ of
  Eq.\ (\ref{eq-total-energy}). Being an eigenstate, there is no need to impose
  the equation of continuity. The latter is automatically satisfied in the
  state $\vert \mathcal{G}\rangle$, $J = J_{q}$, but this ground state cannot
  sustain a finite current, \emph{i.~e.}, $J=0$. One should remark in this
  context that the assumption that all the model parameters of the Hamiltonian
  are real (which allows to consider that all the eigenstates are real)
  excludes superconductivity: the average of the current operator $j_{q}$ of
  Eq.\ (\ref{eq-j}) vanishes for any real state.}

\bibitem[{\citenamefont{Baer and Neuhauser}(2003)}]{BaerNeuhauser:03}
\bibinfo{author}{\bibfnamefont{R.}~\bibnamefont{Baer}} \bibnamefont{and}
  \bibinfo{author}{\bibfnamefont{D.}~\bibnamefont{Neuhauser}},
  \bibinfo{journal}{Chem. Phys. Lett.} \textbf{\bibinfo{volume}{374}},
  \bibinfo{pages}{459 } (\bibinfo{year}{2003}).

\bibitem[{\citenamefont{Frensley}(1991)}]{frensley:91}
\bibinfo{author}{\bibfnamefont{W.~R.} \bibnamefont{Frensley}},
  \bibinfo{journal}{Rev. Mod. Phys.} \textbf{\bibinfo{volume}{63}},
  \bibinfo{pages}{215} (\bibinfo{year}{1991}).

\end{thebibliography}
\end{document}